\documentclass[12pt,a4paper]{article}

\usepackage{cite}
\usepackage{graphicx,amsmath,amssymb}

\DeclareMathAlphabet{\mathpzc}{OT1}{pzc}{m}{it}

\date{}
{ 
\title{ \bf  \Large  Bootstrap \& momentum transfer dependence in small $x$ evolution equations} 

\author{  G. Chachamis$^1$, A. Sabio Vera$^2$, C. Salas$^2$ 
\bigskip \\ { \normalsize
 $^1$ Instituto de F{\' \i}sica Corpuscular, CSIC-U. de Val{\`e}ncia,}\\ {\normalsize Parc Cient{\' \i}fic, E-46980 Paterna, Valencia, Spain.}\\ \normalsize
$^2$ Instituto de F{\' \i}sica Te{\' o}rica UAM/CSIC \& U. Aut{\' o}noma de Madrid, \\  \normalsize   
E-28049 Madrid, Spain
}

}
\begin{document} 

 


\maketitle 

Using Monte Carlo integration techniques, we investigate running coupling effects compatible with the high energy 
bootstrap condition to all orders in the strong coupling in evolution equations valid at small values of Bjorken 
$x$ in deep inelastic scattering. A model for the running of the coupling with analytic behavior in the 
infrared region and compatible with power corrections to jet observables is used.  As a difference to 
the fixed coupling case, where the momentum transfer acts as an effective strong cut-off of the diffusion to infrared scales, 
in our running coupling study the dependence on the momentum transfer is much milder.

\section{Theoretical set up}
In this work we present a study of a long-standing problem in high energy QCD: the treatment of 
the running of the strong coupling in evolution equations driven by the Balitsky-Fadin-Kuraev-Lipatov (BFKL) equation~\cite{BFKL1,BFKL2,BFKL3}. This subject has been extensively discussed in the 
literature (see, {\it e.g.} Refs.~\cite{running}). We put the emphasis on a particular form of writing the equation with running coupling which is consistent, in principle to all orders in a coupling expansion, with the bootstrap property of QCD scattering amplitudes at high energies~\cite{Braun:1994mw,Levin:1994di,Kovchegov:2006vj}. Bootstrap in high energy QCD has been discussed in many papers~\cite{Fadin:1998fv}  
and we refer the reader to the original work by Braun~\cite{Braun:1994mw} for a detailed discussion 
directly related to our present study. Our new contribution is to be able to study the problem using Monte Carlo integration techniques which solve the BFKL equation with a running coupling exactly and allow us to access exclusive information of the final states in the cut amplitude case, and of the diffusion pattern in the virtual diagrams for the non-forward elastic amplitude. We are particularly interested in the dependence of our solution on the total momentum transfer. For a connection of 
this representation of the BFKL equation with renormalon contributions we refer the reader to~\cite{Levin:1994di} and for a more recent related analysis in coordinate space to~\cite{Kovchegov:2006vj}.

\begin{figure}[htbp]
  \centering
  \includegraphics[width=11cm,angle=0]{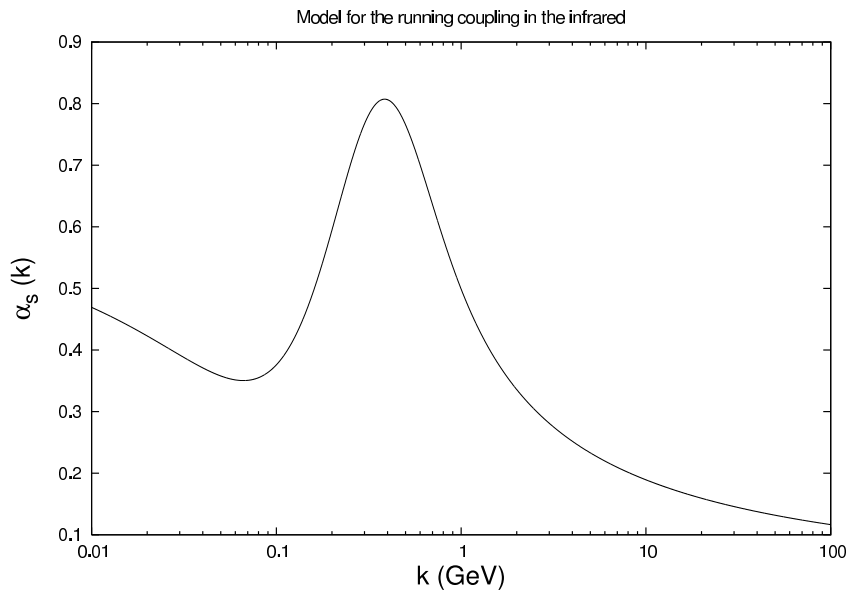}
  \caption{Model for the running of the coupling in the infrared.}
  \label{Alphas}
\end{figure}

To model the running of the coupling in the infrared we will use a simple parametrization 
introduced by Webber in Ref.~\cite{Webber:1998um} which at low momentum scales is 
consistent with global data of infrared power corrections to perturbative observables (mainly 
related to jet event shapes). The relevant formula reads
\begin{eqnarray}
\alpha_s \left(k\right) &=& \frac{4 \pi}{\beta_0} \left(\frac{1}{\ln{\frac{k^2}{\Lambda^2}}}
+\frac{125 \left(\Lambda^2 + 4 k^2\right)}{\left(\Lambda^2 - k^2\right)\left(4 + \frac{k^2}{\Lambda^2}\right)^4}\right),
\label{PCalpha}
\end{eqnarray}
which, for $\beta_0 = (11 N_c -2 n_f)/3$, $n_f=3$ and $\Lambda=0.25\,{\rm GeV}$, gives $\alpha_s \left(91 {\rm GeV}\right) = 0.118$. Its dependence on $k$ is shown in Fig.~\ref{Alphas}.

In the following we introduce the parametrization~(\ref{PCalpha}) of the running of the coupling in a version of the BFKL equation for a projection on a SU(3) singlet in the t-channel which can be directly derived from Eq.~(4.12) in Ref.~\cite{Forshaw:1997dc}\footnote{For a Monte Carlo study of the total momentum transfer dependence of the BFKL gluon Green function in the LO and NLO adjoint representations in QCD and $N=4$ SUSY see Refs.~\cite{Reps}}. If we  use\footnote{In this text all momenta are two-dimensional and we take the notation 
$\vec{p} = p$, for moduli we use $|p|$}
\begin{eqnarray}
\epsilon (q) &=& - \frac{{\bar \alpha}_s}{4 \pi} \int \frac{d^2 k}{k^2}\frac{q^2}{(k-q)^2}
\end{eqnarray}
as the notation for the gluon Regge trajectory, we can then write the non-forward BFKL equation 
with a fixed coupling for the $t$-channel partial wave $f_\omega (k_1,k_2;q)$ as
\begin{eqnarray}
\bigg(\omega - \epsilon(k_1)-\epsilon(k_1-q)\bigg) f_\omega (k_1,k_2;q) ~=~ \delta^{(2)} (k_1-k_2) && 
\nonumber\\ 
&&\hspace{-9.5cm} - \frac{{\bar \alpha}_s}{2 \pi} \int d^2 l \Bigg[\frac{q^2}{(l-q)^2 k_1^2}
- \frac{1}{(l-k_1)^2} \left(1+\frac{(k_1-q)^2 l^2}{(l-q)^2 k_1^2}\right)\Bigg] f_\omega(l,k_2;q),
\label{NFBFKL}
\end{eqnarray}
with ${\bar \alpha}_s \equiv \alpha_s N_c / \pi$.  Note that the solution to this equation corresponds to a four point 
Green function for four off-shell 
reggeized gluons carrying two-dimensional transverse momenta $-k_1, k_1-q, k_2,q-k_2$, all of them outgoing. $q$ corresponds 
to the total momentum transfer in the $t$-channel. 
In order to match the normalization used by Braun in~\cite{Braun:1994mw} it is now needed to introduce the rescaling
\begin{eqnarray}
{\mathcal G}_\omega (l,k_2;q) &\equiv& f_\omega(l,k_2;q) \frac{l^2}{k_1^2}.
\end{eqnarray}
The set up under study can be simply implemented by replacing each of the 
squared transverse momenta $p^2$ of Eq.~(\ref{NFBFKL}) with a general function $\eta (p)$. 
The new trajectory will then be
\begin{eqnarray}
\epsilon (q) &=& - \int \frac{d^2 k}{4 \pi} 
\frac{\eta(q)}{\eta(k) \eta(k-q) } 
\end{eqnarray}
and Eq.~(\ref{NFBFKL}) now reads
\begin{eqnarray}
\bigg(\omega - \epsilon(k_1)-\epsilon(k_1-q)\bigg) {\mathcal G}_\omega (k_1,k_2;q) 
=  \delta^{(2)} (k_1-k_2) && \nonumber\\ 
&&\hspace{-9.5cm} + \int \frac{d^2 l}{2 \pi} \frac{\eta(k_1) }{\eta(l)\eta(k_1-l)}
\Bigg[1+ \frac{\eta(k_1-q) \eta(l) - \eta(q) \eta(k_1-l)}{\eta(l-q) \eta(k_1) }\Bigg] 
{\mathcal G}_\omega(l,k_2;q).
\end{eqnarray}
As it has been shown in~\cite{Braun:1994mw}, this equation is compatible with the bootstrap condition for the all-orders expansion of 
the function $\eta$. 

Let us first study the simplest case, with $q=0$, corresponding to forward scattering, or, by the optical theorem, to a contribution to the total cross section. From now on we also fix $\eta(k) = k^2 / {\bar \alpha}_s(k)$ with $\alpha_s$  as in Eq.~(\ref{PCalpha}). We, therefore, can write
\begin{eqnarray}
\bigg(\omega - 2 \, \epsilon(k_1)\bigg) {\mathcal G}_\omega (k_1,k_2) 
= \delta^{(2)} (k_1-k_2)  + \int \frac{d^2 l}{\pi} \frac{\eta(k_1) }{\eta(l)\eta(k_1-l)}
{\mathcal G}_\omega(l,k_2)
\end{eqnarray}
since $\eta(0) = 0$. 

For our Monte Carlo implementation of the solution to this equation (see Refs.~\cite{Diff} 
for similar studies in the fixed coupling case)  it is convenient to introduce a shift in the integration momentum of the form 
$l= k + k_1$ and  a mass parameter $\lambda$ to 
separate the resolved real emissions (with $k^2 > \lambda^2$) from the unresolved ones (with 
$k^2 < \lambda^2$). The latter, after integration over the phase space of the emitted gluons, generate infrared divergences which should cancel against those of the gluon Regge trajectories. The final results here presented are independent of $\lambda$ in the limit $\lambda \to 0$. Taking the approximation ${\mathcal G}_\omega(k+k_1,k_2) \simeq 
{\mathcal G}_\omega(k_1,k_2)$ for unresolved emissions we can then write 
\begin{eqnarray}
\bigg(\omega - 2 \, \epsilon_\lambda (k_1)\bigg) {\mathcal G}_\omega (k_1,k_2) 
&=&  \delta^{(2)} (k_1-k_2) \nonumber\\
&&\hspace{0.cm} +  \int \frac{d^2 k}{\pi} \frac{\eta(k_1)\theta\left(k^2-\lambda^2\right)}{\eta(k)\eta(k+k_1)}
{\mathcal G}_\omega(k+k_1,k_2),
\end{eqnarray}where
\begin{eqnarray}
\epsilon_\lambda (q) &=& - \int \frac{d^2 k}{2 \pi} 
\frac{\eta(q) \theta\left(k^2-\lambda^2\right)}{\eta(k) \left(\eta(k)+\eta(k-q)\right)}. 
\end{eqnarray}
We go back to Bjorken $x$ space using
\begin{eqnarray}
{\cal F} (k_1,k_2,x) &=& \int \frac{d \omega}{2 \pi i} x^{-\omega} {\mathcal G}_\omega(k_1,k_2). 
\end{eqnarray}
The final expression to be evaluated using Monte Carlo integration techniques is
\begin{eqnarray}
{\cal F} (k_1,k_2,x) = x^{- 2\epsilon_\lambda (k_1)} 
\Bigg\{ \delta^{(2)} \left(k_1-k_2\right) + \sum_{n=1}^\infty \prod_{i=1}^n \int {d^2 p_i \over \pi}
\frac{\eta \left(k_1+ \sum_{l=1}^{i-1} p_l \right)}{\eta(p_i)\eta \left(k_1+ \sum_{l=1}^i p_l \right)} &&
\nonumber\\
&&	\hspace{-14.5cm} \times \theta(p_i^2 - \lambda^2)  \delta^{(2)} \left(k_1+ \sum_{l=1}^n p_l -k_2 \right)  \int^1_{x_{i-1}} \frac{d x_i}{x_i} \, x_i^{2 \left(\epsilon_\lambda \left(k_1+ \sum_{l=1}^{i-1} p_l \right)
-\epsilon_\lambda \left(k_1+ \sum_{l=1}^{i} p_l \right)\right)}\Bigg\},
\label{GGFite}
\end{eqnarray}
where $x_0 \equiv x$.  Note that $n$ corresponds to the number of on-shell gluons emitted with a longitudinal momentum 
fraction $x_i$ and a transverse momentum $k_i$.

\begin{figure}[htbp]
  \centering
  \includegraphics[width=11cm,angle=0]{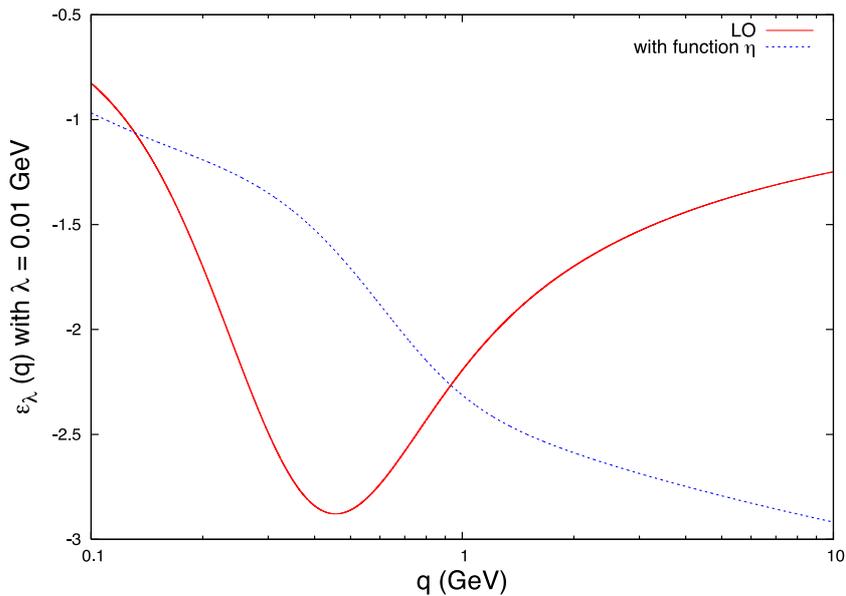}
  \caption{Gluon Regge trajectory in different schemes}
  \label{T-Eta}
\end{figure}

For completeness, we compare the gluon trajectory in the form 
\begin{eqnarray}
\epsilon_\lambda (q) &=& - \frac{q^2}{{\bar \alpha}_s (q)} \int \frac{d^2 k}{2 \pi k^2} \frac{{\bar \alpha}_s(k) {\bar \alpha}_s(k-q)}{k^2+(k-q)^2} \theta(k^2-\lambda^2),
\end{eqnarray}
with the usual one at leading order in our $\lambda$-regularization of infrared divergencies 
\begin{eqnarray}
\epsilon^{\rm LO}_\lambda (q) &=& - {\bar \alpha}_s (q) q^2  \int \frac{d^2 k}{2 \pi k^2} 
\frac{\theta(k^2-\lambda^2)}{k^2+(k-q)^2} 
~\simeq~ - \frac{{\bar \alpha}_s (q)}{2} \ln{\frac{q^2}{\lambda^2}}
\end{eqnarray}
in Fig.~\ref{T-Eta} (both lines are calculated with $\lambda=0.01$ GeV). 
Note that the behaviour of both representations is quite different at large values of the modulus of the 
transverse momentum in the reggeized gluon propagators (which correspond to the power-like terms in Eq.~(\ref{GGFite})). 

A similar iteration of the BFKL kernel applies for the more complicated non-forward equation. 
In this case Eq.~(\ref{GGFite}) should be
\begin{eqnarray}
{\cal F} (k_1,k_2,q,x) &=& x^{- \left(\epsilon_\lambda (k_1)+\epsilon_\lambda (k_1-q)\right)} 
\Bigg\{ \delta^{(2)} \left(k_1-k_2\right) \\
&&	\hspace{-3.cm} + \sum_{n=1}^\infty \prod_{i=1}^n \int {d^2 p_i \over 2 \pi}
\frac{\eta \left(k_1+ \sum_{l=1}^{i-1} p_l \right)}{\eta(p_i)\eta \left(k_1+ \sum_{l=1}^i p_l \right)} 
\theta(p_i^2 - \lambda^2)  \delta^{(2)} \left(k_1+ \sum_{l=1}^n p_l -k_2 \right) \nonumber\\
&&	\hspace{-3.cm} \times 
\Bigg[1+ \frac{\eta \left(k_1+ \sum_{l=1}^{i-1} p_l -q \right) \eta \left(k_1+ \sum_{l=1}^{i} p_l \right) - \eta \left(q \right) 
\eta \left(p_i \right)}{\eta \left(k_1+ \sum_{l=1}^{i} p_l-q \right) 
\eta \left(k_1+ \sum_{l=1}^{i-1} p_l \right) }\Bigg]  \nonumber\\
&&	\hspace{-3.cm} \times \int^1_{x_{i-1}} \frac{d x_i}{x_i} \, x_i^{ 
\left(
\epsilon_\lambda \left(k_1+ \sum_{l=1}^{i-1} p_l \right)
+\epsilon_\lambda \left(k_1+ \sum_{l=1}^{i-1} p_l-q \right)
-\epsilon_\lambda \left(k_1+ \sum_{l=1}^{i} p_l \right)
-\epsilon_\lambda \left(k_1+ \sum_{l=1}^{i} p_l-q \right)\right)}\Bigg\}. \nonumber
\label{GGFiteNF}
\end{eqnarray}
The regularization of infrared divergencies is identical to the forward scattering case, with a unique infrared regulator 
$\lambda$. Clearly the $|q| \to 0$ limit of Eq.~(\ref{GGFiteNF}) corresponds to Eq.~(\ref{GGFite}). We are now ready to present the numerical results for the evaluation of both functions 
in Eqs.~(\ref{GGFite}) and~(\ref{GGFiteNF}).

\section{Numerical results \& discussion}

\begin{figure}[htbp]
  \centering
  \includegraphics[width=11cm,angle=0]{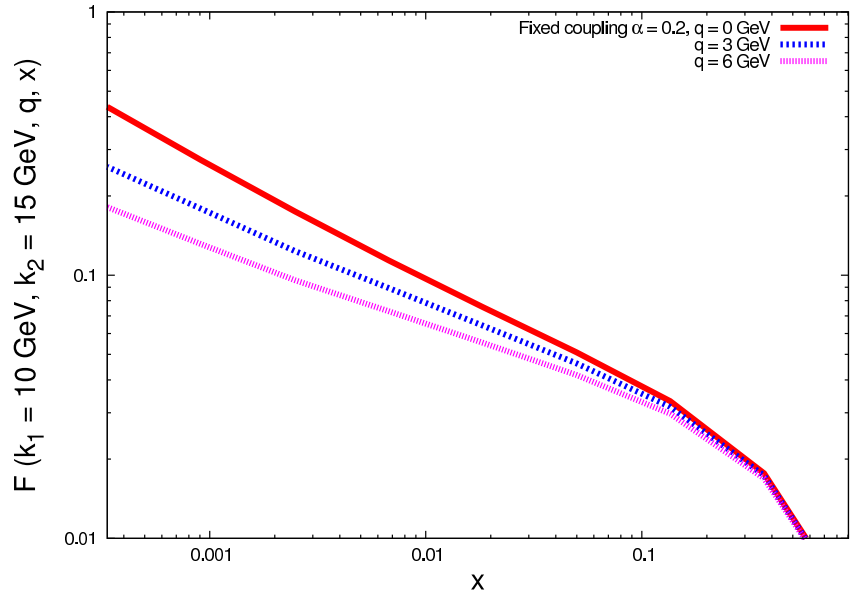}
  \caption{Evolution of the gluon Green function with the Bjorken variable $x$, for fixed values of the transverse momenta and 
  a non-running coupling. }
  \label{xFqFixed}
\end{figure}

We start by investigating the evolution with $x$ of the gluon Green function for fixed coupling and show our results in 
Fig.~\ref{xFqFixed}. In the region of $x$ where we have performed our calculations ($x > 0.0003 $) we observe that 
the growth of the function as $x$ decreases is more pronounced for small $q$. Due to $SL(2,C)$ invariance the asymptotic 
slope of our curves will be the same in the $x \to 0$ limit. We have presented the plot for ${\bar \alpha}_s = 0.2$, 
$|k_1| = 10 \, {\rm GeV}$ and $|k_2| = 15 \,  {\rm GeV}$, but our results are generic. 

\begin{figure}[htbp]
  \centering
  \includegraphics[width=11cm,angle=0]{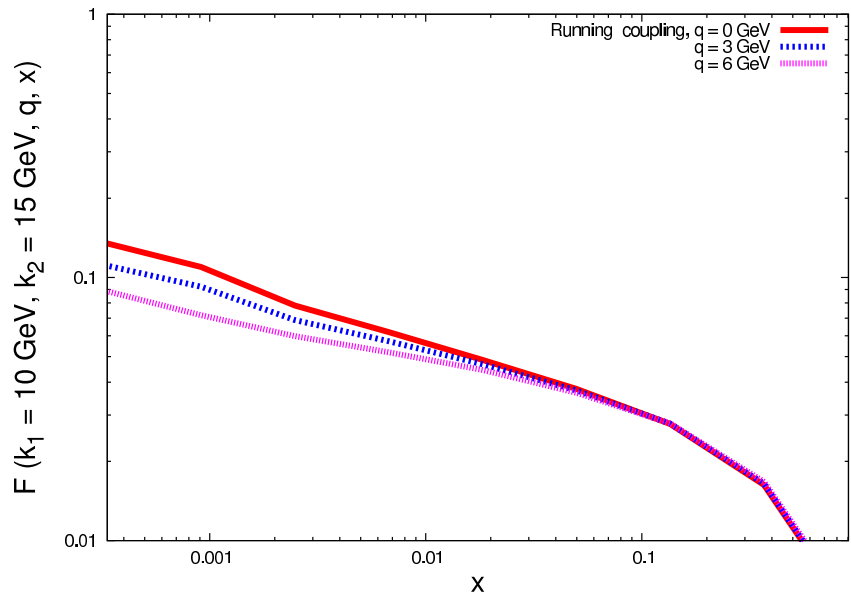}
  \caption{Evolution of the gluon Green function with the Bjorken variable $x$, for fixed values of the transverse momenta and 
  a running coupling. }
  \label{xFqRunning}
\end{figure}

As it is well-known, the effect of introducing the running of the coupling is to reduce the growth of the Green function as $x$ goes to zero. 
The advantage of our method of calculation to previous analysis in the literature is that we can solve the BFKL equation exactly, with no 
asymptotic approximations. In general, we find that the introduction of the running of the coupling in a form compatible with bootstrap to all orders indeed reduces the growth of the solution. As in the fixed coupling case, with running we see in Fig.~\ref{xFqRunning} that we 
have a smaller Green function as the momentum transfer $q$ grows.  However, in the running coupling case we have to go to 
smaller values of $x$ if we want to make this feature manifest. 

\begin{figure}[htbp]
  \centering
  \includegraphics[width=11cm,angle=0]{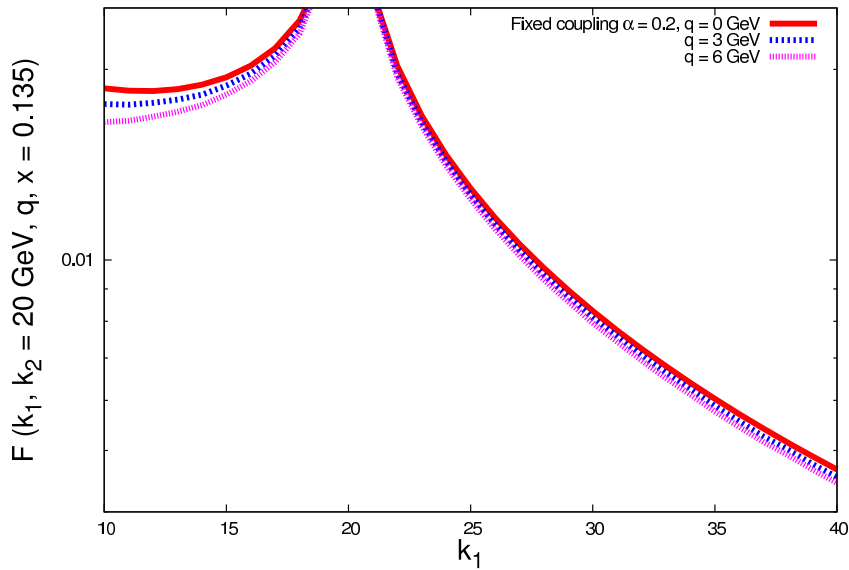}
  \caption{Collinear behaviour of the gluon Green function, for fixed values of one transverse momentum, a non-running coupling 
  and $x=0.135$.}
  \label{xFqFixedCollinearY2}
\end{figure}

\begin{figure}[htbp]
  \centering
  \includegraphics[width=11cm,angle=0]{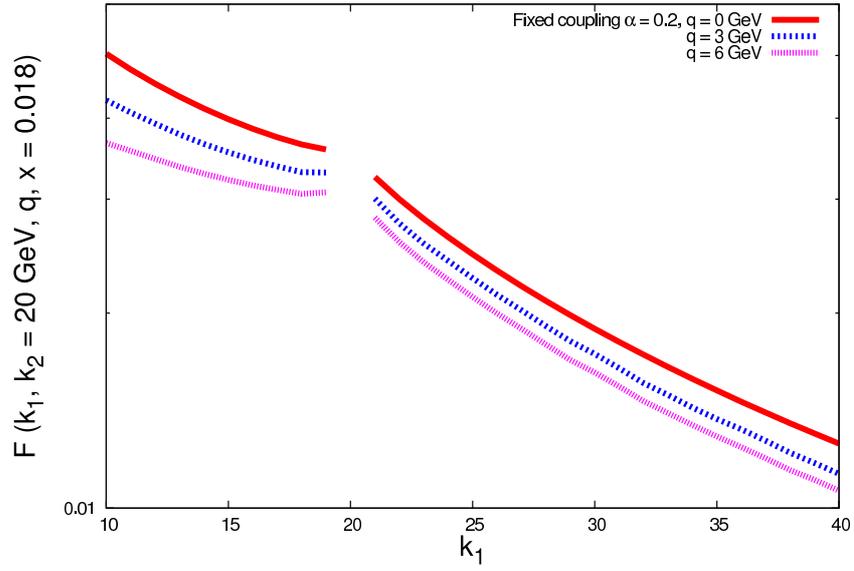}
  \caption{Collinear behaviour of the gluon Green function, for fixed values of one transverse momentum, a non-running coupling 
  and $x=0.018$.}
  \label{xFqFixedCollinearY4}
\end{figure}

In order to investigate this effect in more detail we can perform an analysis 
of the gluon Green function in the collinear regions where $|k_1| \gg |k_2|$ and vice verse. Let us fix $|k_2| = 20 \,  {\rm GeV}$, and 
study the solution to our equation for fixed and running couplings, with different values of $x$. In Fig.~\ref{xFqFixedCollinearY2}, for a 
fixed coupling and a large value of $x$ ($x=0.135$) it can be seen that the effect of increasing the value of $q$ is important only in the 
region $|k_1| \ll |k_2|$. However, when we go to higher energies and decrease $x$, as in Fig.~\ref{xFqFixedCollinearY4}, 
we see that the effect of $q$ is manifest in the full 
range of $|k_1|$.

\begin{figure}[htbp]
  \centering
  \includegraphics[width=11cm,angle=0]{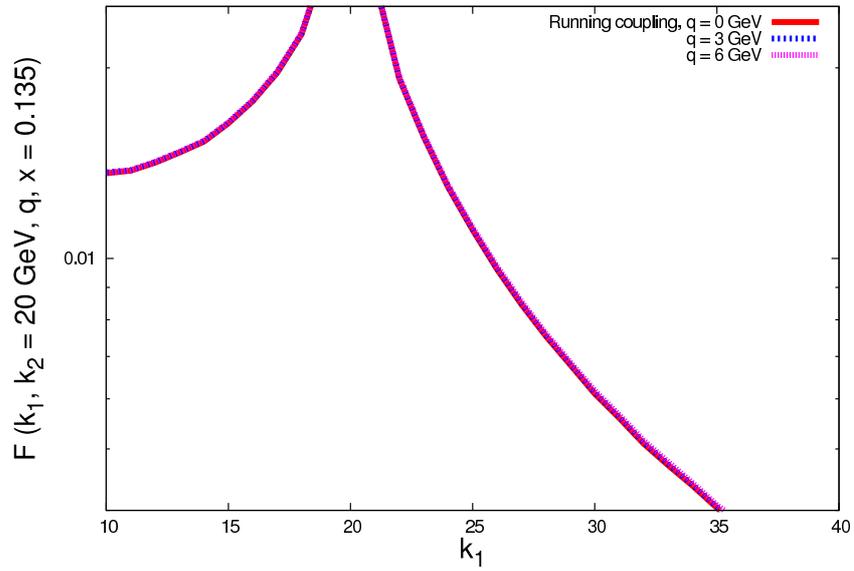}
  \caption{Collinear behaviour of the gluon Green function, for fixed values of one transverse momentum, a running coupling 
  and $x=0.135$.}
  \label{xFqRunningCollinearY2}
\end{figure}

\begin{figure}[htbp]
  \centering
  \includegraphics[width=11cm,angle=0]{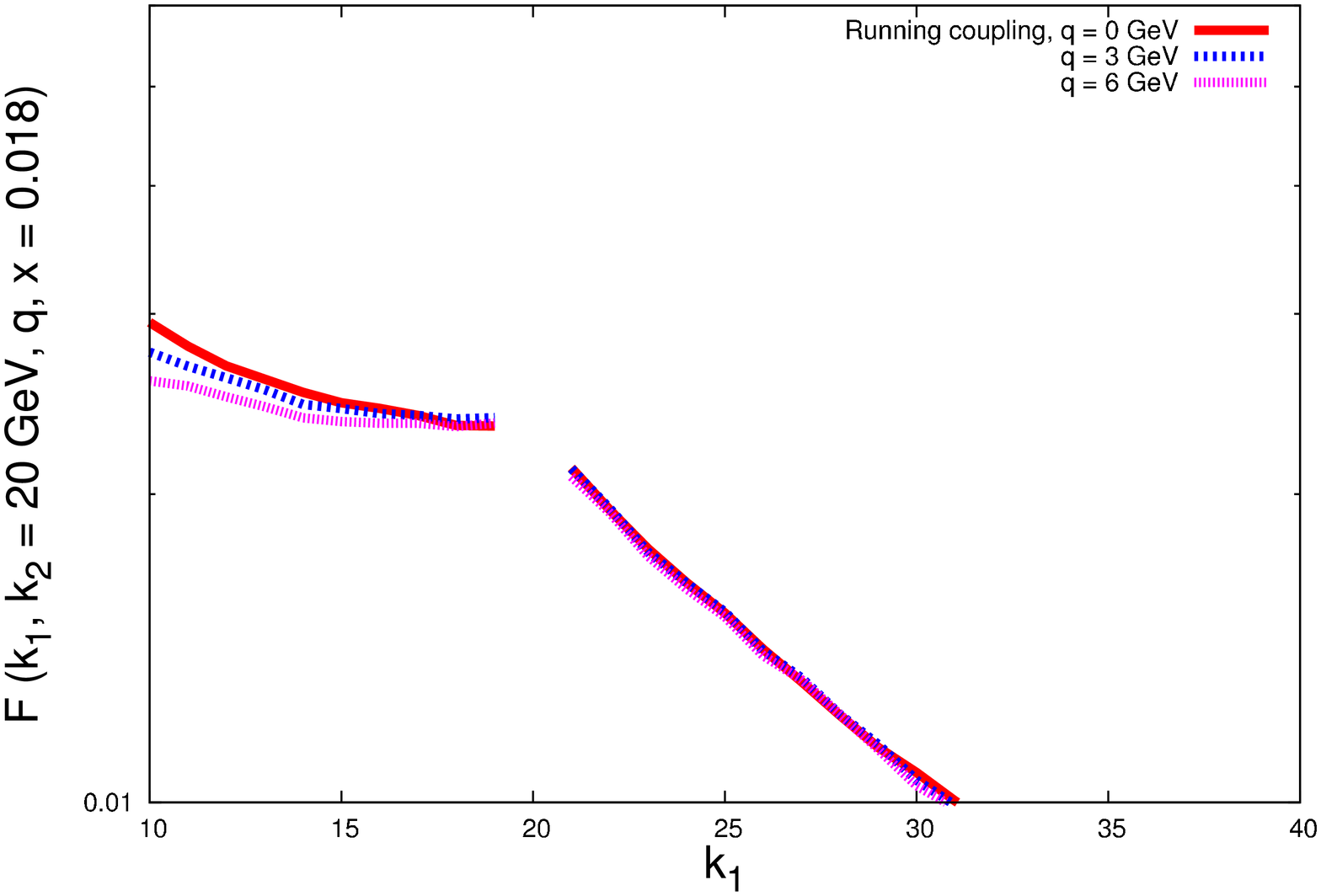}
  \caption{Collinear behaviour of the gluon Green function, for fixed values of one transverse momentum, a running coupling 
  and $x=0.018$.}
  \label{xFqRunningCollinearY4}
\end{figure}

It is striking that the running of the coupling completely eliminates any effect of introducing the momentum 
transfer for $x=0.135$ and $q$ from zero up to 6 GeV. This can be seen in Fig.~\ref{xFqRunningCollinearY2}.
It is needed to reduce the value of $x$ by one order of magnitude in order to start finding a decrease of the Green function as $q$ 
increases, for $|k_1| < |k_2|$, as in the fixed coupling case. This is shown in Fig.~\ref{xFqRunningCollinearY4}.

\begin{figure}[htbp]
  \centering
  \includegraphics[width=11cm,angle=0]{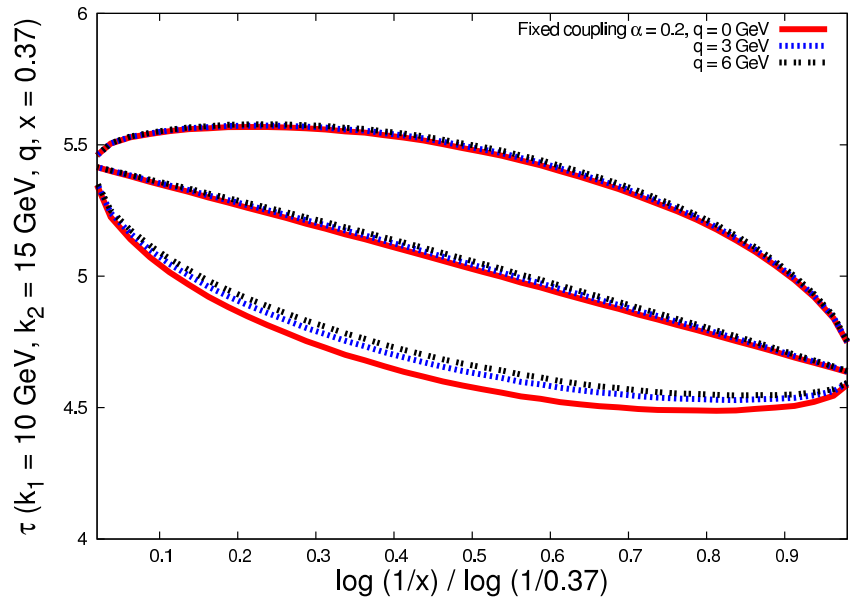}
  \caption{Distribution of the transverse momenta in the internal propagator of the gluon ladder for a fixed coupling and a large 
  value of Bjorken $x$.}
  \label{DiffusionqY1-Fixed}
\end{figure}

\begin{figure}[htbp]
  \centering
  \includegraphics[width=11cm,angle=0]{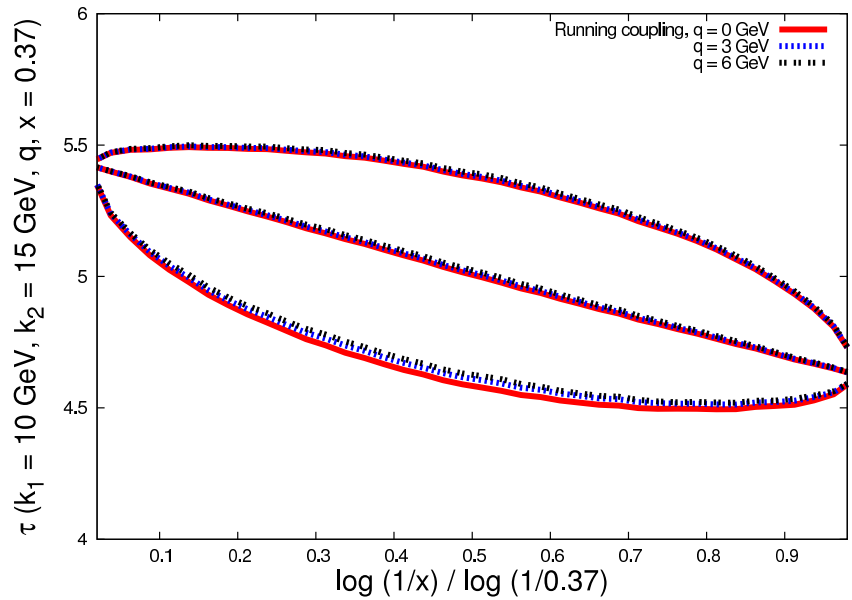}
  \caption{Distribution of the transverse momenta in the internal propagator of the gluon ladder for a running coupling and a large 
  value of Bjorken $x$.}
  \label{DiffusionqY1-Running}
\end{figure}

Since we have full access to the exclusive information of all momenta configurations 
in the gluon ladder let us investigate the typical transverse momentum scale 
running in the internal propagators. This corresponds to the well-known diffusion ``cigar"~\cite{Bartels:1995yk} where 
the mean value of the variable $\tau = \log{<p_i^2>/({\rm GeV}^2)}$ is plotted (together with the lines of one standard deviation towards    
the infrared and ultraviolet for the gluons in the BFKL ladder) as a function of the normalized rapidities of the corresponding gluon lines. For ${\bar \alpha}_s=0.2$ in Fig.~\ref{DiffusionqY1-Fixed} we show the typical effect of introducing a non-zero momentum transfer: the diffusion to the 
infrared is reduced as $q$ increases, while the ultraviolet diffusion line remains stable. In other words, the momentum transfer 
acts as an effective infrared cut-off. 
For the same large value of $x=0.37$ we find in Fig.~\ref{DiffusionqY1-Running} a smaller suppression of the infrared diffusion 
in the set up with a running coupling as the momentum transfer increases. We also see that the diffusion to the ultraviolet is 
suppressed with respect to the fixed coupling case, independently of the value of $q$.

\begin{figure}[htbp]
  \centering
  \includegraphics[width=11cm,angle=0]{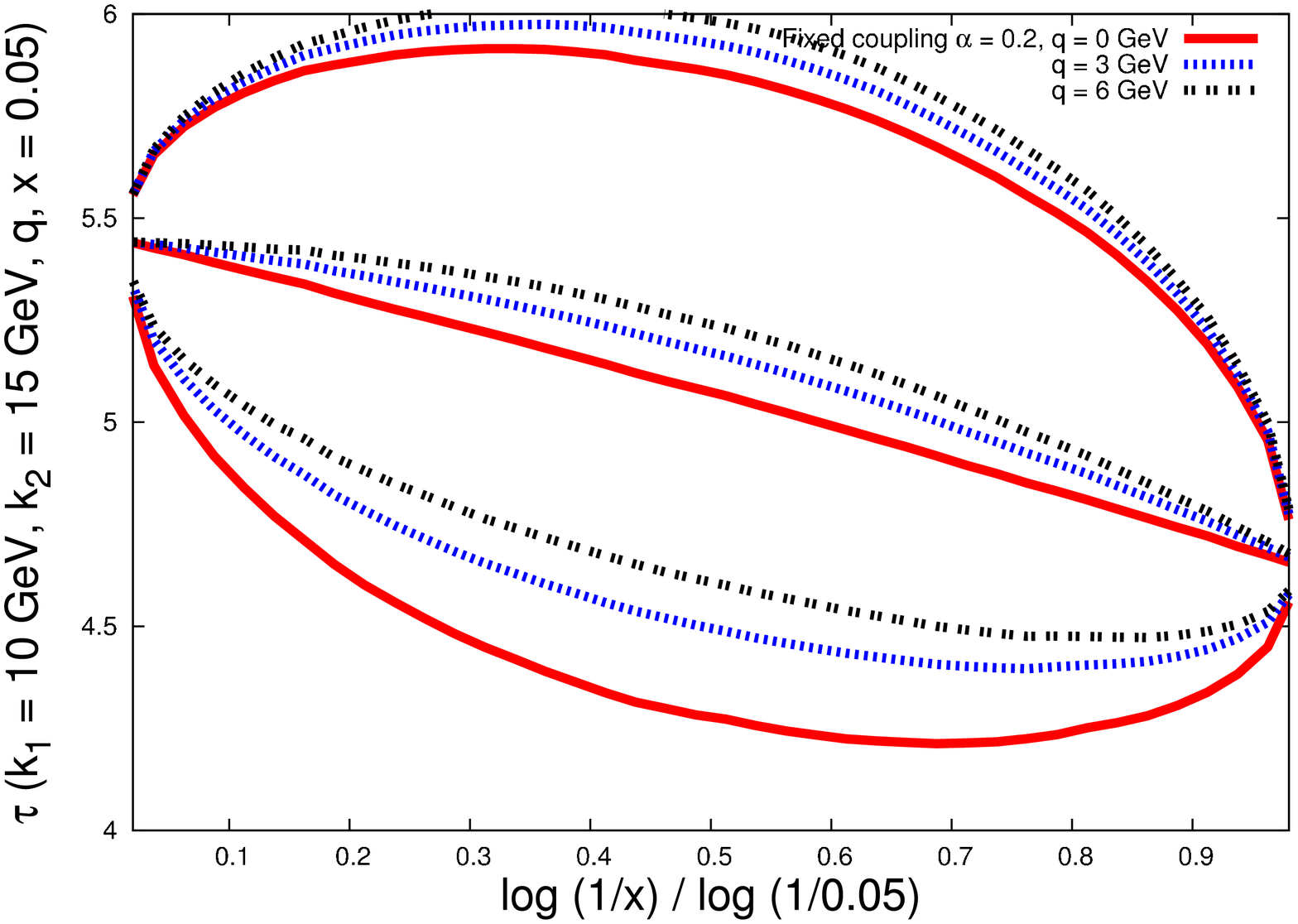}
  \caption{Distribution of the transverse momenta in the internal propagator of the gluon ladder for a fixed coupling and a small  
  value of Bjorken $x$.}
  \label{DiffusionqY3-Fixed}
\end{figure}

\begin{figure}[htbp]
  \centering
  \includegraphics[width=11cm,angle=0]{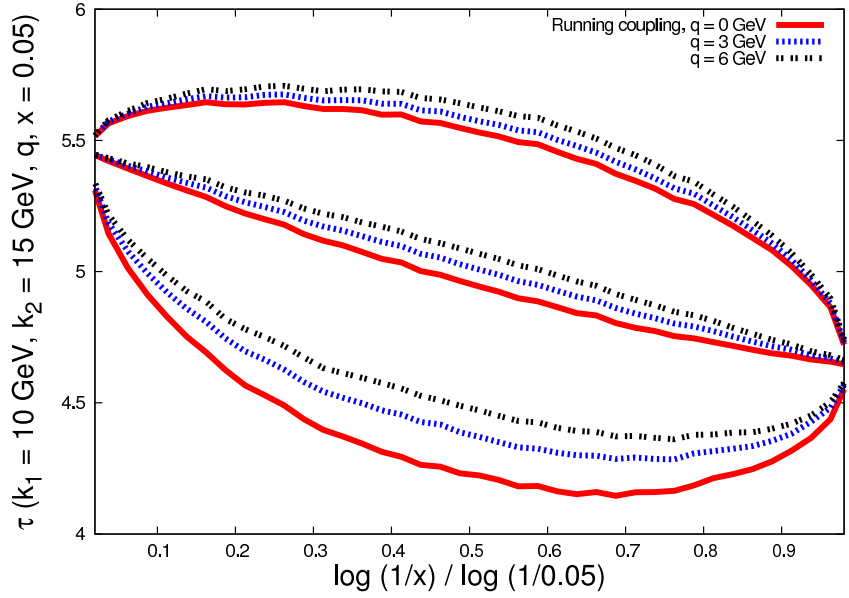}
  \caption{Distribution of the transverse momenta in the internal propagator of the gluon ladder for a running coupling and a small  
  value of Bjorken $x$.}
  \label{DiffusionqY3-Running}
\end{figure}

For a smaller value of $x$ the spread in transverse momentum in the internal gluon propagators is much bigger, in particular 
for a fixed coupling, see 
Fig.~\ref{DiffusionqY3-Fixed}. The modification of the diffusion picture depending on the values of $q$ is interesting since the 
suppression of the evolution into the infrared is very big but it is also present with the opposite effect, an enhancement, in the ultraviolet region. In the latter a large 
momentum transfer pushes the gluon momenta to live in more perturbative regions of phase space. 
This diffusion to the ultraviolet is somewhat reduced in the case with a running coupling, see Fig.~\ref{DiffusionqY3-Running}. 
As the value of $x$ gets smaller we can see that the influence of introducing a non-zero momentum transfer is larger, always 
``pulling" the diffusion ``cigar" towards more perturbative regions in both,  the fixed and running coupling scenarios.

\section{Conclusions \& outlook}
We have investigated small $x$ evolution equations which incorporate a running coupling compatible with bootstrap 
to all orders in a perturbative expansion. Our analysis is exact since we have found the solution to the BFKL equation 
using an iteration in transverse momentum space which is finally expressed in terms of integrals over transverse momenta and the rapidity of the 
internal gluon propagators, which we have evaluated using Monte Carlo integration techniques.  We have found that the 
effect of the total momentum transfer as an effective cut-off for diffusion of transverse momenta into the infrared region is very 
suppressed in this way of running the strong coupling. The propagation of the gluon ladders into ultraviolet regions is also 
suppressed, independently of the value of the momentum transfer. Our next task will be to integrate these results with suitable 
impact factors in order to gauge the phenomenological relevance of the results here presented. 
$~$
\\
\\
\\
{\bf \large Acknowledgements}\\
G. C. thanks the Paul Scherrer Institut, the Department of Theoretical Physics at the Aut{\'o}noma University of Madrid and the ``Instituto de F{\'\i}sica Te{\' o}rica UAM / CSIC" for their hospitality. Research partially supported by the European Comission under contract LHCPhenoNet (PITN-GA-2010-264564), 
the Comunidad de Madrid through Proyecto HEPHACOS ESP-1473, and MICINN (FPA-2010-17747).

\end{document}